# MRI lung lobe segmentation in pediatric cystic fibrosis patients using a recurrent neural network trained with publicly accessible CT datasets

**Authors:** Orso Pusterla[1,2,3], Rahel Heule[4,5], Francesco Santini[1,3,6], Thomas Weikert[6], Corin Willers[2], Simon Andermatt[3], Robin Sandkühler[3], Sylvia Nyilas[7], Philipp Latzin[2], Oliver Bieri[1,3], and Grzegorz Bauman[1,3]

**Author Affiliations**

[1]*Department of Radiology, Division of Radiological Physics, University Hospital Basel, University of Basel, Basel, Switzerland*

[2]*Division of Pediatric Respiratory Medicine and Allergology, Department of Pediatrics, Inselspital, Bern University Hospital, University of Bern, Switzerland*

[3]*Department of Biomedical Engineering, University of Basel, Basel, Switzerland*

[4]*High Field Magnetic Resonance, Max Planck Institute for Biological Cybernetics, Tübingen, Germany*

[5]*Department of Biomedical Magnetic Resonance, University of Tübingen, Tübingen, Germany*

[6]*Department of Radiology, University Hospital Basel, University of Basel, Basel, Switzerland*

[7]*Department of Diagnostic, Interventional and Pediatric Radiology, Inselspital, Bern University Hospital, University of Bern, Switzerland*

**Corresponding author:**

Orso Pusterla, PhD

Division of Radiological Physics, University Hospital Basel, Petersgraben 4, 4031 Basel, Switzerland

E-mail: orso.pusterla@unibas.ch

**Original submission accepted by Magnetic Resonance in Medicine as Research Article.**

**Paper details:** Abstract 250 words (max 250). Paper Body 5000 words (max 5000), 9 Figures and 1 Table (max 10). References 52.

**Supplementary Material** for review and online publication: 1 Appendix, 1 Table, 3 Supporting Figures.

**Running head title:** RNN MRI lung lobe segmentation with CT data.

**Acknowledgements:** Orso Pusterla acknowledges the support of the Swiss Cystic Fibrosis Society (CFCH).

**ABSTRACT**

**Purpose**

To introduce a widely applicable workflow for pulmonary lobe segmentation of MR images using a recurrent neural network (RNN) trained with chest computed tomography (CT) datasets. The feasibility is demonstrated for 2D coronal ultra-fast balanced steady-state free precession (ufSSFP) MRI.

**Methods**

Lung lobes of 250 publicly accessible CT datasets of adults were segmented with an open-source CT-specific algorithm. To match 2D ufSSFP MRI data of pediatric patients, both CT data and segmentations were translated into pseudo-MR images, masked to suppress anatomy outside the lung.

Network-1 was trained with pseudo-MR images and lobe segmentations, and applied to 1000 masked ufSSFP images to predict lobe segmentations. These outputs were directly used as targets to train Network-2 and Network-3 with non-masked ufSSFP data as inputs, and an additional whole-lung mask as input for Network-2. Network predictions were compared to reference manual lobe segmentations of ufSSFP data in twenty pediatric cystic fibrosis patients. Manual lobe segmentations were performed by splitting available whole-lung segmentations into lobes.

**Results**

Network-1 was able to segment the lobes of ufSSFP images, and Network-2 and Network-3 further increased segmentation accuracy and robustness. The average all-lobe Dice similarity coefficients were 95.0±2.8 (mean±pooled SD [%]), 96.4±2.5, 93.0±2.0, and the average median Hausdorff distances were 6.1±0.9 (mean±SD [mm]), 5.3±1.1, 7.1±1.3, for Network-1, Network-2, and Network-3, respectively.

**Conclusions**

RNN lung lobe segmentation of 2D ufSSFP imaging is feasible, in good agreement with manual segmentations. The proposed workflow might provide access to automated lobe segmentations for various lung MRI examinations and quantitative analyses.

## INTRODUCTION

Pulmonary diseases are one of the most significant public health challenges, severely reducing life quality and expectancy.[1] The current clinical gold standards for morphological and functional assessment of the lung are computed tomography (CT) and nuclear medicine imaging techniques.[2] However, these imaging modalities pose the risk of ionizing radiation. This becomes concerning and contra-indicated especially for longitudinal monitoring of patients, pediatric subjects, and pregnant women.[3,4] Magnetic resonance imaging (MRI) of the lung offers a valuable alternative[5–9] while not involving any potentially harmful ionizing radiation,[10] making it well suited for frequent and long-term examinations.[11–17] Consequently, pulmonary MRI has potential to accelerate diagnosis and treatment monitoring.[11,16,18,19]

For an improved disease evaluation and early diagnosis, scientists and clinicians have recently been focusing on quantifying several pulmonary MR biomarkers, e.g., impaired ventilation and perfusion, structural scoring, and tissue relaxometry.[20–27] These pulmonary biomarkers often require delineating lung tissue boundaries for disease quantification. Manual organ segmentation is time-consuming, but lately, several groups successfully developed algorithms for automated whole-lung segmentation, which appears to be a solved challenge,[28–34] in particular thanks to advances in artificial neural network (ANN) designs. In the context of medical image segmentation, ANNs proved to outperform traditional algorithms in terms of accuracy.[35] In addition, besides providing automated data analyses, ANNs reduce the inter- and intra-observer variability compared to manual segmentations[28], consequently improving the sensitivity in detecting subtle longitudinal clinical changes.

It is expected that biomarkers quantification on the lung lobar level[27,33] provides an improved regional analysis. Moreover, it might provide an increased sensitivity and specificity compared to a whole lung analysis since the lung lobes can be regarded as independent functional units. In fact, the lobes are invaginated and divided by visceral pleurae, as well as supplied by different airways and vessels which do not extend from one lobe to another. Pathologic changes in the lung might be predominantly located in specific lobes. Therefore, lobar segmentation might increase the sensitivity of the biomarkers, consequently improving longitudinal evaluations as well as the disease characterization and phenotyping, ultimately allowing for personalized treatment selection and planning. From a clinical perspective, automated lobar segmentation would likely help radiologists to review the regional involvement of the disease more efficiently.

To date, lobar segmentation of MR data is still hardly accessible since the fissures separating the lobes are most of the time not directly visible or only faintly discernible. Due to the difficulty in obtaining lung lobe segmentations, MRI biomarker quantification on a lobar level is yet barely explored. In some clinical studies,[36,37] patients underwent both CT as well as MR imaging in consecutive sessions, and the segmentation of MR lung lobes (and even lung segments[36,37]) was efficiently obtained by deformable registration of CT lobe masks, computed with CT-specific algorithms. However, the requirement of both CT and MR data is not ideal and a direct approach for lobar segmentation of MR data appears beneficial.

To the best of the author's knowledge, only three studies focused on developing an algorithm for direct lobe segmentation of MR data.[30,33,38] Tustison et al.[30] were the first to propose an atlas-based probability model resulting from manually labeled CT datasets applied to 3D MRI data. Nevertheless, in their method, only the lung shapes are considered for lobe segmentation and no image features such as lung texture or vessel locations are taken into account. Lee et al.[38] introduced a combined multi-atlas and machine learning segmentation model; but it requires matched CT and 3D MRI data acquired in the same patients for training. Winther et al.[33] recently demonstrated the feasibility of lobe segmentation specifically for 3D T1-weighted dynamic-contrast-enhanced (DCE) perfusion imaging by using a supervised convolutional neural network (CNN) trained with manually labeled images.

Pulmonary MR lobe segmentation has thus potential to be automated by the aid of supervised ANNs. Nevertheless, the application of supervised ANNs for lung lobe segmentation and other tasks (e.g., vessel segmentation or detection of pathologies) is still generally hampered by the required large amount of labeled training datasets, which is often not available. Furthermore, specific radiological investigations frequently include MRI with different sequences and contrasts.[13,14] Consequently, supervised ANN training would necessitate labeled datasets for every image contrast (e.g., an ANN trained on T1-weighted data might not perform well on T2-weighted data).

To overcome these limitations, in this study, we propose a widely applicable workflow for MR lung lobe segmentation using a supervised recurrent neural network (RNN[39,40]) primarily trained with publicly available CT data and lobe masks, rather than MR images and manually labeled MR ground truth lobe masks. In fact, the lung tissue on both CT and MR data appears similar, i.e., hyperdense/hyperintense vessel structures are surrounded by hypodense/hypointense parenchyma tissue, as illustrated in Figure 1. A neural network could thus learn pulmonary tissue features from CT data –such as the expected anatomical location of fissures, vessels structures and vessel rarefication next to fissures– and predict results on lung MR data. On the other hand, regions outside the lung exhibit strongly different image features and signal intensities on MR and CT data (cf. Figure 1), and therefore must be masked out. Without masking, the ANN would learn features from CT data, which do not resemble the MR data, likely resulting in reduced performance and erroneous results when applied to predict segmentations of MR data. To increase the similarity between MR and CT data further, we introduce a lung-specific CT-to-MR data translation to generate pseudo-MR data. As a prerequisite for the proposed workflow, MRI lobe segmentation requires an available whole-lung segmentation of MR data to mask out non-lung tissues.

This study investigates the feasibility of obtaining lung lobe segmentation for 2D ultra-fast balanced steady-state free precession (ufSSFP) MRI data. The ufSSFP acquisitions are generally used for free-breathing and contrast agent-free ventilation as well as perfusion imaging with matrix pencil (MP) decomposition,[41,42] which has shown promising results.[17,28,43,44] A lobar quantification of pulmonary functions might ultimately provide valuable imaging biomarkers and a novel dimension for secondary analyses.

**METHODS**

<u>Unrolled workflow</u>

A schematic of the study workflow is presented in Figure 2. Publicly accessible chest CT data[45] of 250 adult patients were used (Figure 2A) and the lung lobes were segmented with the open-source CNN of Hofmanninger et al.[46] To increase the similarity to the 2D coronal ufSSFP data, the 3D CT data and lobe masks were reformatted to 2D coronal "pseudo-MR" (see the subsection "CT data and lobe masks translation into pseudo-MR" hereafter). The first RNN (Network-1) was trained with masked pseudo-MR data as input, i.e., the background regions outside the lung were set to zero, and multi-class lobe masks as targets. These targets were obtained from the pseudo-MR lobe segmentations generated by the open-source CT algorithm (see subsection "CT data and lobe masks translation into pseudo-MR"). The feasibility of 2D pseudo-MR lobe segmentation via Network-1 was assessed through independent pseudo-MR testing data not included into training (Figure 2A).

One thousand coronal 2D ufSSFP images of cystic fibrosis (CF) pediatric patients were included in the study to further optimize the RNNs for ufSSFP imaging lobe segmentation (Figure 2B). Network-1 was applied to predict the lung lobes of 1000 masked and normalized ufSSFP data. The whole-lung segmentations to mask the ufSSFP data were derived with the neural network previously presented by Pusterla et al.[29] The 1000 lung lobe masks predicted by Network-1 served then as targets for training Network-2 and Network-3. Inputs for Network-2 are non-masked ufSSFP data and whole-lung segmentations, while inputs for Network-3 are only non-masked ufSSFP data. To note, non-masked MRI data were used as input for Network-2 and Network-3 to let anatomical image features of the complete thorax be learned by the RNN (e.g., rib cage, spine, heart). Neither selection nor refinements of the data predicted by Network-1 were made in order to evaluate a workflow without user interaction.

MRI data of 20 CF pediatric patients providing 144 2D ufSSFP independent testing images (not contained in the training process) were included in the study to assess the performances of Network-1, Network-2, and Network-3 against reference manual lobe segmentation (Figure 2C). The manual lobe segmentations were performed by drawing the fissure locations on available whole-lung segmentations obtained with the network of Pusterla et al.[29] To note, manual lobe segmentations and those predicted with Network-1 and Network-2 rely on a priori knowledge of the lung shape as they include masked data or whole-lung segmentations as inputs to remove the bias of different whole-lung segmentation boundaries (cf. Willers et al.[28]) on the network performance comparison. On the other hand, we did not incorporate any a priori knowledge about the lung shape into Network-3 to evaluate a fully independent process and allow a fairer comparison to literature results. Nevertheless, to provide additionally an unbiased comparison of Network-3 against Network-1 and Network-2, a consensus whole-lung segmentation defined as the intersection of the whole-lung masks between Network-1 and Network-3 after lobe mask prediction is derived and used for quantitative analysis.

CT chest data and CT lung lobe segmentations

Volumetric chest CT scans from the LUNA16 dataset[45] (available at https://luna16.grand-challenge.org) were used in our study. The LUNA16 dataset, a subset of the LIDC-IDRI database,[47] includes 888 examinations with axial slice thickness ≤ 2.5 mm and pulmonary nodules annotations for developing automatic nodules detection algorithms. In our work, we included patients with lung nodules of diameter ≤ 8 mm, totaling 250 CT chest examinations. The average CT resolution was 0.7±0.1 × 0.7±0.1 × 1.3±0.6 $mm^3$ (mean± standard deviation; anterior-posterior, right-left, head-feet axis, respectively) with a matrix size of 512±0 × 512±0 × 294±134.

The lung lobes on CT data were segmented with the CNN of Hofmanninger et al.[46] (available at https://github.com/JoHof/lungmask), well-established for whole-lung segmentation (DSC = 97%, achieved within the LObe and Lung Analysis 2011 challenge, https://lola11.grand-challenge.org ) and yielding promising preliminary results for lung lobe segmentations. The lungs were segmented into left upper (LU), left lower (LL), right upper (RU), right middle (RM), and right lower (RL) lobes. The lung nodules were retained in the lobe segmentations. All obtained segmentations were included in our study and no manual refinements were made. Representative CT lung lobe segmentations obtained with the Hofmanninger algorithm are presented in Figure 3.

CT data and lobe masks translation into pseudo-MR

Volumetric CT of adult patients were reformatted to pseudo-MR to match 2D coronal ufSSFP MRI data of pediatric patients as close as possible. To this end, the following image processing steps were performed:

1) While ufSSFP data for different patients are acquired with a constant coronal field-of-view (FOV), and a constant matrix-size of 256×256 (and thus constant in-plane resolution), the FOV and matrix-size (and correspondingly the resolution) of the LUNA16 CT data vary. Moreover, the average lung size of the adult patients in the CT cohort differs from the average lung size of the pediatric subjects in the MR cohort. Accordingly, the coronal FOV of every CT dataset was adjusted considering the average coronal lung area, and the in-plane matrix-size was downsampled to 256×256 (cubic interpolation). This process ensured identical coronal matrix size and similar resolution as well as matching average lung sizes of pseudo-MR and MRI data.

2) Chest CT data are acquired at submillimeter resolution and the FWHM of the point spread function (PSF) of the reconstructed images is lower than the PSF of ufSSFP; i.e., CT lung images appear sharper than MR images (cf. Figure 1). Moreover, the visual sharpness of MR images varies among subjects. Hence, to increase the resemblance of the pseudo-MR to MRI, the data were filtered with three different Gaussian filters with kernel sizes of 5x5 and σ ( standard deviation) of 0.75, 1, and 1.25. The Gaussian filter parameters were chosen empirically, by visually estimating the similarity between pseudo-MR and MR data.

3) The slice thickness of pseudo-MR must be adjusted to mimic the ufSSFP data, which were acquired with a slice thickness of 12mm at full width at half maximum (FWHM). The anterior-to-posterior lung size of our MR pediatric population strongly differs between subjects and from the CT adult population. Therefore, a 12mm slice might cover a thicker portion of the lung in a young subject as compared to an adult. Computer simulations of the actual MR slice profile were performed, and coronal CT slices were summed accordingly to simulate pseudo-MR data with three different slice thicknesses: 12mm, 18mm, and 24mm at FWHM.

4) In a last step, the pseudo-MR datasets were normalized. Since the pulmonary signal intensity histograms are positively skewed, the lung signal intensities were logarithmically scaled and then normalized.

Pseudo-MR data in which the lung was not visible were discarded. Pseudo-MR with three different Gaussian filters and three different slice thicknesses were simulated, for a total of approximately 500'000 datasets (250 patients*222 slices*3 filters*3 thicknesses).

The process to reformat the 3D CT lobe masks to 2D coronal pseudo-MR lobe masks was largely analogous to generating pseudo-MR images:

1) FOV adjustment and down-sampling (nearest neighbor interpolation).

2) CT coronal lobe mask combination to simulate the three different slice thicknesses (12, 18, 24 mm). The lobe masks were concatenated and weighted according to the MR slice profile. On a voxel basis, the most frequent value (the mode) occurring between the concatenated lobe masks, defined the reformatted lobe segmentation.

3) The lobe masks were smoothed with a median filter (kernel size 7x7 in coronal orientation).

Figure 4 displays representative pseudo-MR images and lobe masks along with 2D ufSSFP MR images for comparison.

<u>Pseudo-MR cohort splitting for RNN processing</u>

The pseudo-MR patient cohort was randomly divided into 178, 35, and 37 patients for RNN training, validation, and testing, respectively. The subsets included approximately 350’000, 75’000, and 75’000 datasets (corresponding to 70%, 15%, and 15%), respectively.

MRI study population and data acquisition

All MRI data included in this work are retrospectively obtained from partly published single-center observational studies performed at the University Children's Hospital of Bern, Switzerland, on a 1.5T whole-body MRI Scanner (MAGNETOM Aera; Siemens Healthineers, Erlangen, Germany).[17,28,44] The children underwent a comprehensive MRI study protocol for morphological and functional pulmonary assessment without contrast agent. The children were not sedated during the scans. All MR measurements at the Children’s Hospital of Bern and all data processing within this study were conducted in agreement with the local ethic regulations. Written informed consent was obtained by parents and by participants if older than 14 years.

In this methodological study, MR data from 100 children with CF participating in the aforementioned clinical studies were used. Some subjects underwent MR imaging more than once at different investigation time-points, and in total, 147 examinations were performed. For functional assessment of the whole lung with ufSSFP, coronal images were acquired at 6-11 anterior-posterior slice positions. Hence, overall, 1144 two-dimensional ufSSFP datasets were included in our study. On average, ufSSFP imaging was acquired at 8 slice positions per examination (mean 7.8, standard deviation 1.4 slices).

The ufSSFP data were acquired for ventilation and perfusion functional imaging computed with matrix pencil decomposition MRI (MP-MRI).[42] Two-dimensional time-resolved ufSSFP coronal sets were acquired during 48 seconds of free-breathing.[41] Scan parameters for ufSSFP were as follows: field-of-view = 425mm×425mm, matrix size = 128×128 interpolated to 256×256 (in-plane resolution = 1.66mm×1.66mm), slice thickness = 12mm, echo time / repetition time (TE/TR) = 0.67/1.46 ms, bandwidth = 2056 Hz/pixel, flip angle = 65°, GRAPPA factor 2, acquisition rate = 3.3 images/s, 160 coronal images per slice. In our study, only one coronal ufSSFP image per slice, i.e., one time point, was included.

MR cohort splitting for RNN processing

The MR patient cohort, which included 100 patients and a total of 147 examinations, was divided into 65, 15, and 20 patients for RNN training, validation, and testing (corresponding to 112, 15, and 20 MR examinations), respectively. The validation and testing cohort was randomly selected based on children who underwent only one MR examination. The training, validation, and testing subsets included overall 900, 100, and 144 images (approximately 78%, 9%, and 13%), respectively. None of the subjects of the testing dataset was included in the RNN training or validation. The testing data represent thus a new and independent set for neural network testing purposes.

The principal patient demographics for the training and validation were as follows [mean±SD, if not stated otherwise]: age 12.2±3.5 [years] (range 5.6 to 18.9); 42% males; weight 43.6±16.0 [kg]; height 1.51±0.19 [m]; forced expiratory volume in 1s: -1.42±1.57 [z-score]; lung clearance index at 2.5%: 10.8±3.3 [turnovers].

For the testing cohort, the patient demographics were: age 10.4±3.7 [years] (range 4.7 to 17.2); 45% males; weight 42.0±16.4 [kg]; height 1.46±0.17 [m]; forced expiratory volume in 1s: -1.28±1.13 [z-score]; lung clearance index at 2.5%: 11.1±3.1 [turnovers].

MR data postprocessing and manual lobe segmentation

The whole lung of all the ufSSFP data was segmented with a previously trained RNN as presented by Pusterla et al. and Willers et al.[28,29] Masked ufSSFP data were calculated by multiplying the ufSSFP images by the binary whole-lung segmentation masks.

To compare the lobe masks predicted by the artificial neural networks, the lung lobes of 20 patients in the testing dataset (totaling 144 ufSSFP images) were manually segmented by an MR-scientist with seven years of experience in thoracic imaging (O.P.). The lobe segmentations were performed by splitting the available whole-lung segmentations of 2D ufSSFP data, i.e., the lung boundaries were fixed. The lobe divisions were identified based on the coronal 2D ufSSFP data and other MRI datasets (e.g., sagittal and axial views of 3D ufSSFP). The inter-slice lobe segmentation consistency in ufSSFP images was also verified. A chest radiologist with four years of experience (T.W.) reviewed and refined the lobe masks, if necessary. The segmentations were drawn using MITK.[48]

Multi-dimensional gated recurrent units neural network

A neural network with main layers consisting of multi-dimensional gated recurrent units (MD-GRU) was used for voxel-wise multi-class classification,[39,40] i.e., lung lobe segmentation. During training, to increase the prediction robustness, an on-the-fly data augmentation was applied, i.e., images and masks were randomly and slightly scaled, rotated, skewed, distorted, and shifted, and Gaussian noise was added. More details about the MD-GRU can be found in Andermatt et al.[39,40] and the network is provided at https://github.com/zubata88/mdgru. The MD-GRU has already shown competitive accuracy for brain segmentation tasks.[39,40] Moreover, for whole-lung segmentation, the MD-GRU previously reached a Dice similarity coefficient of 0.93 and performed comparably well as human observers.[28,29]

A description of the input, target, and predicted output data for Network-1, Network-2, and Network-3 is provided in the "Unrolled workflow" section and summarized in the Supporting Information "Input, target, and predicted data of Network-1, Network-2, and Network-3" available online. All the networks in this study were trained from scratch, and no pre-trained models were used. The computations were performed on a mid-range working station with a GPU (NVIDIA Quadro P6000, NVIDIA Corp. Santa Clara, CA).

Data analysis and statistics

To quantify the agreement between two lobe segmentation masks ($M_1$ and $M_2$), the Sørensen Dice similarity coefficient (DSC) was calculated:[49]

$\mathrm{DSC}(M_1, M_2) := 2 \cdot \frac{M_1 \cap M_2}{M_1 + M_2}$ [Eq. 1].

The DSC ranges between [0, 1], where zero indicates no overlap and 1 indicates exact overlap of the two masks. Additionally, the lobes volume percentage (LVP) in respect to the whole-lung volume (100%) were calculated and statistics was computed with Pearson's correlation.

Since the workflow is investigated for MR data generally acquired at several slice positions for whole-lung functional assessment, the metrics above were calculated for the whole-lung volumes, which included on average 8 coronal slices per patient. For evaluating pseudo-MR data lobe predictions for the whole lung, to further mimic the MR examinations, eight equidistant coronal slices were chosen to calculate the metrics. The indices were summarized as group mean ± standard deviation (SD).

To evaluate the largest error between two ufSSFP lobe segmentation masks in length, the Hausdorff distance (HD) was calculated for every slice as follows:[49]

$\mathrm{HD}(B_1, B_2) := \max\left( \max_{\vec{x} \in B_1} \min_{\vec{y} \in B_2} \|\vec{x} - \vec{y}\|_2 , \max_{\vec{y} \in B_2} \min_{\vec{x} \in B_1} \|\vec{x} - \vec{y}\|_2 \right)$ [Eq. 2] ,

where $\vec{x}$ and $\vec{y}$ are two points on the boundaries $B_1$ and $B_2$ of the segmentation masks $M_1$ and $M_2$.

The HD is a commonly used performance measure for segmentation accuracy and corresponds to the maximum among all the distances from point ($\vec{x}$) on the boundary ($B_1$) to its closest point ($\vec{y}$) on the boundary ($B_2$). The HD ranges between [0, ∞], where zero indicates total mask overlap. If the HD is not computable, we set HD=∞ (e.g., in case one lobe was segmented on the reference mask but not on the mask to be evaluated, or vice versa). We thus report the median and third quartile (q3/4) among all the coronal slices.

As global parameters, the all-lobe DSC and HD were calculated. We report median, standard deviation, and pooled standard deviation (square root of the average variance) among all the lobes and metrics, if applicable.

For a qualitative assessment, the automatically segmented lobe masks with Network-1, Network-2, and Network-3, were visually evaluated (O.P.) in the CF testing cohort, determining the percentage of masks not well performed that would require further refinements for secondary analyses. The procedure was repeated at three time points for reliability assessment.

**RESULTS**

Pseudo-MR lung lobe segmentation with Network-1

Network-1 proves the feasibility of accurately predicting lobe segmentations of 2D data. On the 37-patient pseudo-MR testing cohort, the DSCs between ground truth and predicted masks were 98.2±1.5, 97.5±2.9, 97.2±1.3, 91.5±4.1, 97.6±1.3 (mean±SD [%]) for left upper, left lower, right upper, right middle and right lower lobes, respectively. The all-lobes averaged DSC was 96.4±2.9 (2.5) {mean±SD (pooled SD) [%]}. Representative pseudo-MR images, ground-truth pseudo-MR masks, and predicted masks are presented in Supporting Information Figure S1 available online.

We analyzed the lobe predictions separately for each of the three slice thicknesses and filters used. These parameters influence the lobe predictions only very marginally, i.e., the DSCs vary of about 0.1%. All respective DSCs are given in the Supporting Information Table S1 available online.

MR lobe predictions via Network-1 and generation of training dataset

Network-1 demonstrated capability to predict the lung lobe with masked-ufSSFP input data. Representative results are presented in Figure 5, and additional examples are shown in Supporting Information Figure S2 available online. The lobes appear reliably segmented. All the 1000 predicted lobe segmentations were used to train Network-2 and Network-3, without any manual mask refinements or data selection.

Manual lobe segmentation and Network-1, Network-2, and Network-3 predictions on the CF testing cohort

The manual lobe segmentation of the ufSSFP testing datasets from 20 patients took about 15 hours including the radiologist's supervision. We estimate a processing time of about 45 minutes per patient (~8 slices). It was time-consuming to locate the fissures based on morphological sequences such as 3D ufSSFP images in sagittal views with corresponding maximum intensity projections by following the vessel trees and then reliably and consistently porting the lobe boundaries onto the 2D ufSSFP images. The large slice thickness of 12mm limits the spatial resolution in 2D ufSSFP and results in partial-volume effects. Therefore, the fissures are not directly visible, and an exact delineation based on 2D ufSSFP data was not possible. We estimate the location of the actual fissures, i.e., the lobar boundaries, to be approximately within ±1 cm from the drawn one.

Multi-class lung lobe segmentation proved to be feasible with all three networks. A lung mask prediction could be executed efficiently in less than two seconds. The training process of a neural network was completed in less than 24 hours.

Comparisons of manual lobe segmentations with the one predicted via the three networks are exemplarily presented in Figure 6, demonstrating good visual agreement. The DSCs, HDs, LVPs, and volumes Pearson's correlations for all the lobes are given in Table 1, and the DSC distributions are presented via violin-plots in Figure 7.

The DSCs, HDs, and LVPs demonstrate a high similarity, accuracy, and agreement between the manual and the predicted lobe masks. The average all-lobes DSCs were 95.0±3.7 (2.8), 96.4±2.7 (1. 5), 93.0±2.8 (2.0) {mean±SD (pooled SD) [%]}, for Network-1, Network-2, and Network-3, respectively. The lung boundaries were not kept fixed for Network-3. Therefore, to objectively compare it to the other networks, the consensus lung mask must be used, yielding an average all-lobes DSC of the manual masks against Network-3 of 96.0±2.9 (1.7) {mean±SD (pooled SD) [%]}. This demonstrates that Network-3 performs in fact comparably well as Network-2. On the other hand, the DSCs of Network-1 and Network-2 are unchanged since the reference manual lobe masks share exactly the same whole-lung boundaries; an illustrative explanatory example is provided in Supporting Information Figure S3 available online. The DSC between the whole-lung binary mask of Network-3 (all lobe classes set to one) and the one of Network-1 (corresponding to the one of Network-2 and the manual lobe segmentation) was 96.8±0.6 (mean±SD [%]).

The average median of the all-lobes HDs were 6.1±0.9 (11.7), 5.3±1.1 (9.4), and 7.1±1.3 (11.5) {mean±SD (mean q3/4) [mm]}, for Network-1, Network-2, and Network-3 against manual segmentations, respectively (see Table 1). The average HDs indicate a good accuracy between the predictions of the neural networks and the manual masks.

The lung lobe volumes for all the segmentations derived from the networks were very highly correlated to the manual one (cf. Table 1) and statistical significance was reached ($r>0.90$, $p<10^{-7}$).

Qualitatively, 88±1%, 93±1%, and 91±1% (mean ± SD) of the lobe masks were well predicted for Network-1, Network-2, and Network-3, respectively. The residual lobe-mask predictions would require minor refinements due to the mixing of lobe boundaries, as representatively shown in Figure 8.

A whole-lung ufSSFP examination of a CF subject along with manual and Network-2 predicted lobe segmentations is illustrated in Figure 9. The predictions of the network show an excellent inter-slice lobe mask consistency. The inferred lobe segmentations with Network-2 have a good similarity with the manual masks as corroborated by the DSCs, HDs, and LVPs.

**DISCUSSION**

Currently, pulmonary MR biomarkers are rarely quantified on a lobar basis since manual segmentations are prohibitively time-consuming and complex to be performed. Supervised ANNs can automatically predict segmentations, but the initial preparation of several datasets requires a considerable amount of manual work by experts. To overcome these limitations, in this study, we introduced a widely applicable ANN-driven workflow that provides effective lung lobe segmentation of MR images without using either ground truth MR labeled datasets or a multitude of MR data (Network-1). The novel method is built from accessible CT datasets of adults, an open-source algorithm to predict CT lobe masks, and lung-specific MR-to-CT image translation to train the first ANN. As a prerequisite, the MR data necessitates whole-lung masking for lobe prediction (Network-1). We further improved the ANN performance and robustness (Network-2 and Network-3) using 1000 MR datasets as input and the lobe segmentations predicted by Network-1 as target. Overall, we also demonstrated the feasibility of accurate lobe segmentation on 2D coronal imaging and for a pediatric CF patient cohort.

Our final predicted lobe masks were evaluated with the DSC, which takes into account the number of correctly and misclassified voxels, and with the HD, which considers the largest errors of deviating spatial boundaries. With respect to ground truth, our network reaches a 93.0±2.8 [%] DSC (all-lobes average) agreement comparable to literature results: 93.4±2.8 [%][33] and 94.6±1.4 [%][38]. However, a direct inter-study comparison is hardly feasible since different methods and datasets were used, and our approach was based on 2D imaging comprising only a limited number of slices per patient. Our ANN achieves an average HD of about 7mm, corroborating the accuracy of our lobe segmentations. We expect that lobe boundaries manually drawn by different thoracic experts would exhibit comparable uncertainties since the lung fissures are not visible in 2D ufSSFP imaging; no exact delineation is possible, and moreover, partial volume effects on the actual lobe shapes further hinder the achievable precision. Although our neural networks deliver satisfactory lobe masks in most of the ufSSFP images, it has to be noted that similarly to many networks for segmentation tasks trained with a limited number of datasets, inaccuracies can occur, and manual segmentation refinements might be required (cf. Figure 8).

Our study has potential for further improvements. Our workflow is based on publicly accessible CT data and lobe segmentations predicted by the Hofmanninger CNN,[46] rather than on ground truth segmentations drawn by radiologists. The lobe segmentations were very well predicted, but minor misclassified areas might occur. It might be beneficial to begin the proposed workflow with ground truth segmentations refined by a radiologist, but such data are hardly obtainable.

The pseudo-MR data generation could be further ameliorated and more datasets could be generated. The volumetric CT could have been pitched, rolled, yawed, skewed, and distorted before pseudo-MR generation. Furthermore, instead of or complementary to using the proposed CT-to-MR translation procedure, deep learning-based imaging synthesis[50] could be adopted to translate chest CT images into MR images, possibly eliminating the need for whole-lung segmentations (cf. input of Network-1). Similarly, an approach based on cross-modality domain adaptation[51] could be investigated to directly adapt the CT-specific algorithm of Hofmanninger et al.[46] to segment MRI data. Besides the difficulty of implementing such techniques, they might require several CT and MR data that are often unavailable. Our proposed approach has the advantage that it does not necessitate initial MR datasets as input for training (Network-1). The workflow requires an available whole-lung segmentation, but this is in fact a prerequisite for all pulmonary quantification processes. Moreover, nowadays, segmentation of relatively simple organs such as the whole lung might be feasible with only a limited number of datasets by using specific ANNs.[34,52] Similarly, unsupervised ANNs might be investigated for MR lung lobe segmentation.

Neither data selection nor mask refinements were performed in this study to develop a novel workflow without need for user interaction. MR lobe masks predicted via Network-1 are employed to train Network-2 and Network-3. Segmentation flaws and inaccuracies of these “silver-standard” datasets might adversely impact the final accuracy of these networks. Other options to possibly augment the performance of our trained networks consist in the use of both MR and CT data as input for RNN training, or in the concurrent processing of all the slices of a patient, stacked for enhanced intra-slice consistency. In addition, in our MR lobe database predicted with Network-1, we included ufSSFP images at one time point only. It appears thus as a straightforward extension to use more time points for improved robustness. Further, modified RNN or CNN architectures (e.g., U-Net, Res-Net) can be investigated.

In this work, we randomly chose the patients for the MR testing cohort. We note that a larger cohort could have been selected and stratified into clinically healthy, moderately, and severely diseased subjects; moreover, the application in other disease groups needs to be shown. Since the experimental lobe segmentation module is already integrated into our fully automated postprocessing pipeline for perfusion and ventilation assessment with MP-MRI, a broader evaluation to assess the clinical potential of quantifying lung lobe functions and structures is warranted in future studies.

The proposed cross-modality, cross-pathology, and cross-age approach might be relevant to foster further MRI pulmonary research, especially in children, in which CT is preferably avoided for frequent examinations. We expect that the proposed workflow for lobe segmentation is broadly applicable and can easily be adapted to quantify pulmonary functions or tissue characteristics acquired with sequences providing diverse contrasts, such as 3D gradient-echo for DCE-MRI, 3D ufSSFP, 3D ultrashort echo time (UTE), or 2D gradient echo (GRE), inversion-recovery imaging, etc. Moreover, we hypothesize that our method might be adaptable to segment the lung vessels or other pathologies (e.g., nodules, embolisms) on MR data, starting from CT labeled datasets and MR whole-lung masks as suggested in this study. In general, ANN-driven CT-to-MR image translation has excellent prospects to automatize pulmonary MR data analysis and to accelerate the clinical inclusion of lung MRI, potentially improving respiratory health care.

## CONCLUSIONS

Lung lobe segmentation of 2D ufSSFP pediatric MR data is feasible with an ANN trained from CT data and in good agreement with respect to ground truth labels. The proposed workflow might provide access to lobe segmentations for various MR lung examinations and secondary analyses.

## DATA AVAILABILITY

The "LUNA" volumetric CT data (https://luna16.grand-challenge.org), the CNN of Hofmanninger et al. for chest CT lobe segmentation (https://github.com/JoHof/lungmask), and the MDGRU ANN of Andermatt et al. (https://github.com/zubata88/mdgru) are available online. The CF patient data are confidential. Upon request, trained networks can be made available from the corresponding author for collaborations.

**TABLES**

| | | **LU lobe** | **LL lobe** | **RU lobe** | **RM lobe** | **RL lobe** | **All-lobes** |
|---|---|---|---|---|---|---|---|
| **Dice similarity coeff. [%]** | | | | | | | **Mean ± SD (pooled SD)** |
| **Network-1** | Mean ± SD | 97.0 ± 1.2 | 96.7 ± 1.3 | 96.6 ± 1.4 | 88.3 ± 5.6 | 96.3 ± 2.1 | 95.0 ± 3.7 (2.8) |
| **Network-2** | Mean ± SD | 97.9 ± 0.7 | 97.6 ± 0.7 | 97.4 ± 0.8 | 91.5 ± 3.0 | 97.5 ± 0.8 | 96.4 ± 2.7 (1.5) |
| **Network-3** | Mean ± SD | 94.5 ± 1.0 | 93.7 ± 1.3 | 94.2 ± 1.4 | 88.0 ± 3.7 | 94.7 ± 1.4 | 93.0 ± 2.8 (2.0) |
| **Hausdorff distance [mm]** | | | | | | | **Mean ± SD of median and q3/4** |
| **Network-1** | Median, q3/4 | 6.8, 12.4 | 6.8, 13.4 | 5.2, 8.5 | 5.0, 12.4 | 6.6, 12.1 | 6.1 ± 0.9, 11.7 ± 1.7 |
| **Network-2** | Median, q3/4 | 6.6, 10.3 | 6.0, 10.3 | 4.8, 6.7 | 3.7, 10.0 | 5.2, 10.1 | 5.3 ± 1.1, 9.4 ±1.4 |
| **Network-3** | Median, q3/4 | 8.5, 13.8 | 8.3, 12.4 | 6.0, 8,7 | 5.6, 11.1 | 7.2, 11.7 | 7.1 ± 1.3, 11.5 ±1.7 |
| **Lobe volumes [% of whole-lung volume]** | | | | | | | **Pooled SD** |
| **Manual** | Mean ± SD | 23.9 ± 2.5 | 21.5 ± 2.5 | 19.5 ± 1.8 | 9.6 ± 2.0 | 25.6 ± 2.7 | 2.3 |
| **Network-1** | Mean ± SD | 23.3 ± 2.5 | 22.0 ± 2.6 | 19.3 ± 2.0 | 9.4 ± 2.2 | 26.1 ± 3.1 | 2.5 |
| **Network-2** | Mean ± SD | 23.9 ± 2.5 | 21.4 ± 2.7 | 19.3 ± 1.8 | 9.2 ± 2.1 | 26.2 ± 2.8 | 2.4 |
| **Network-3** | Mean ± SD | 23.9 ± 2.5 | 21.5 ± 2.6 | 19.4 ± 1.8 | 9.5 ± 1.9 | 25.6 ± 2.7 | 2.3 |
| **Lobe volumes [Pearson's correlation (r)]** | | | | | | | **Mean** |
| **Network-1** | | 0.975 | 0.976 | 0.928 | 0.901 | 0.939 | 0.944 |
| **Network-2** | | 0.983 | 0.985 | 0.951 | 0.914 | 0.965 | 0.960 |
| **Network-3** | | 0.969 | 0.977 | 0.943 | 0.902 | 0.953 | 0.947 |

**Table 1.** Lung lobe segmentation performances assessed by the Dice similarity coefficient and the Hausdorff distance for Network-1, Network-2, and Network-3 against manual segmentations in the MRI testing CF cohort. Additionally, the calculated lung lobe volumes and Pearson's correlation coefficient are reported. The DSC for the right middle lobe is lower than the other lobes partly due to its reduced size (cf. Eq.1). Abbreviations: SD, standard deviation; q3/4, third quartile; LU, left upper lobe; LL, left lower lobe; RU, right upper lobe; RM, right middle lobe; RL, right lower lobe.

**FIGURES**

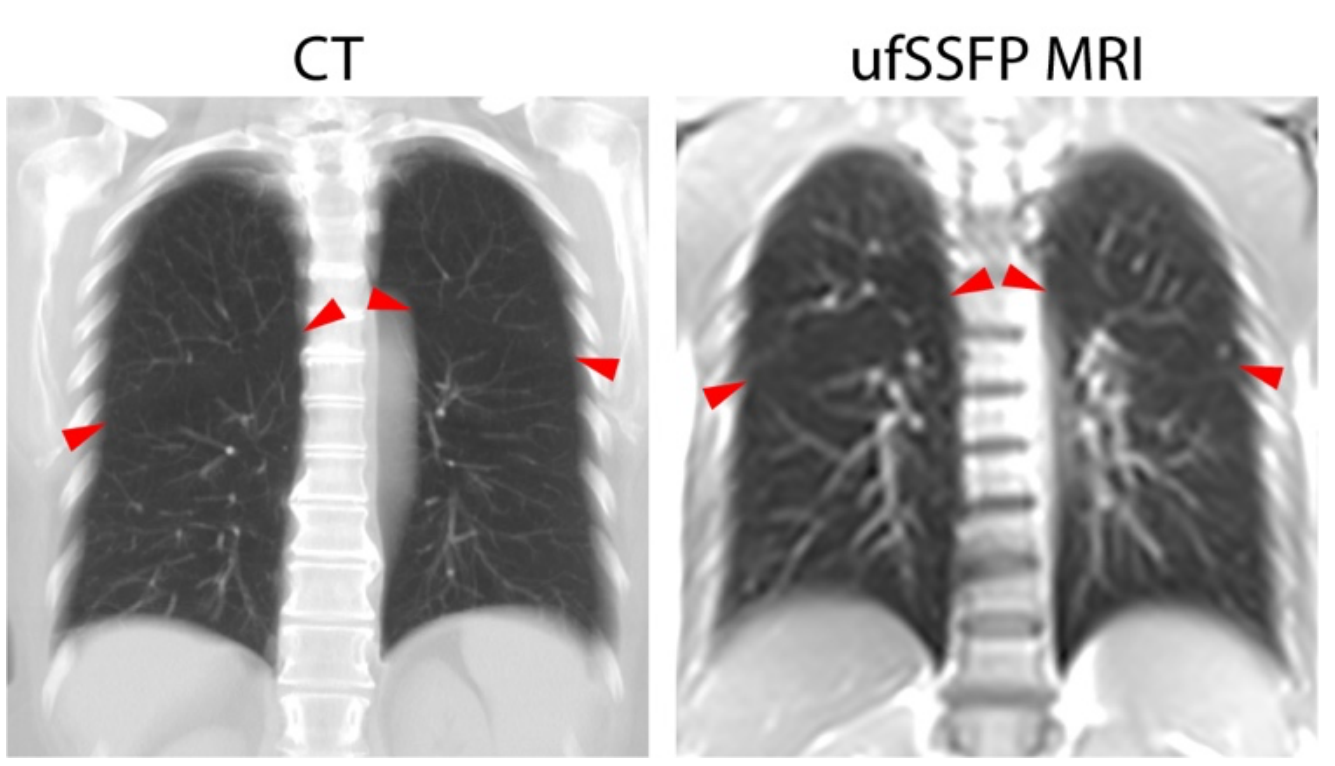

**Figure 1.** Chest CT (left, maximum intensity projection across 15 mm) and MRI (right, 2D ufSSFP with a slice thickness of 12 mm) of two subjects. The lung tissue on both CT and MRI exhibits similar features, i.e., bright vessels overlay the darker pulmonary parenchyma. On the other hand, the bones, the fat, the cartilages, and other tissues have very different contrasts. The red arrowheads indicate the locations of the oblique fissures, which are faintly discernible, separating the upper and lower lobes.

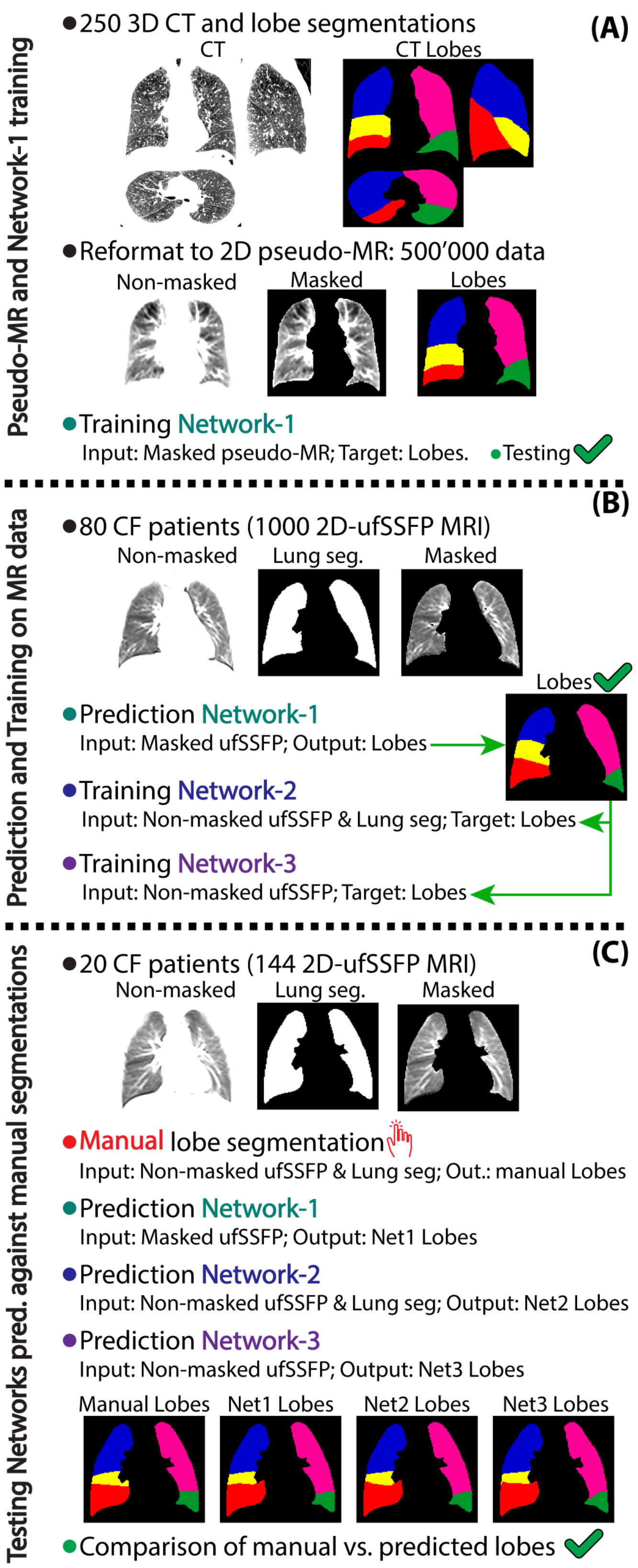

**Figure 2.** Overview of the workflow as investigated in this study. Note: the grayscale-windowing of the CT and MR images is optimized to enhance the visualization of the lung whereas regions outside the lung appear overshoot (i.e., white). The description of the workflow is included in the "*Methods"* section.

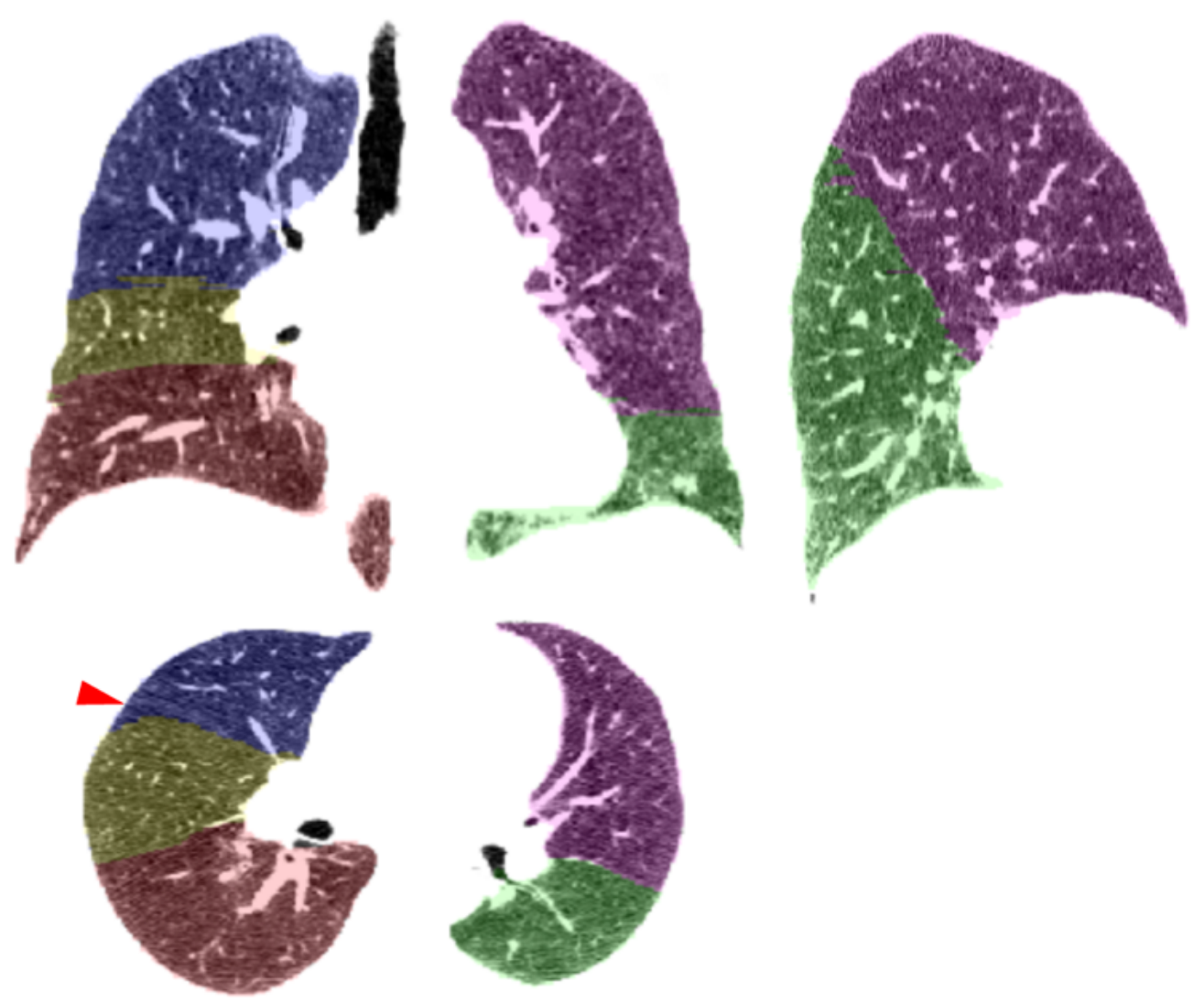

**Figure 3.** Representative lung lobe segmentations predicted with the algorithm developed by Hofmanninger et al. in a CT dataset of the LUNA16 database. The lobe masks are overlaid on the CT base images. The algorithm shows an impressive ability to segment the lung lobes, which are in general well predicted. There are very minor flaws in few datasets, e.g., the fissures are not precisely delineated by few pixels, as indicated by the red arrowhead (axial reformat).

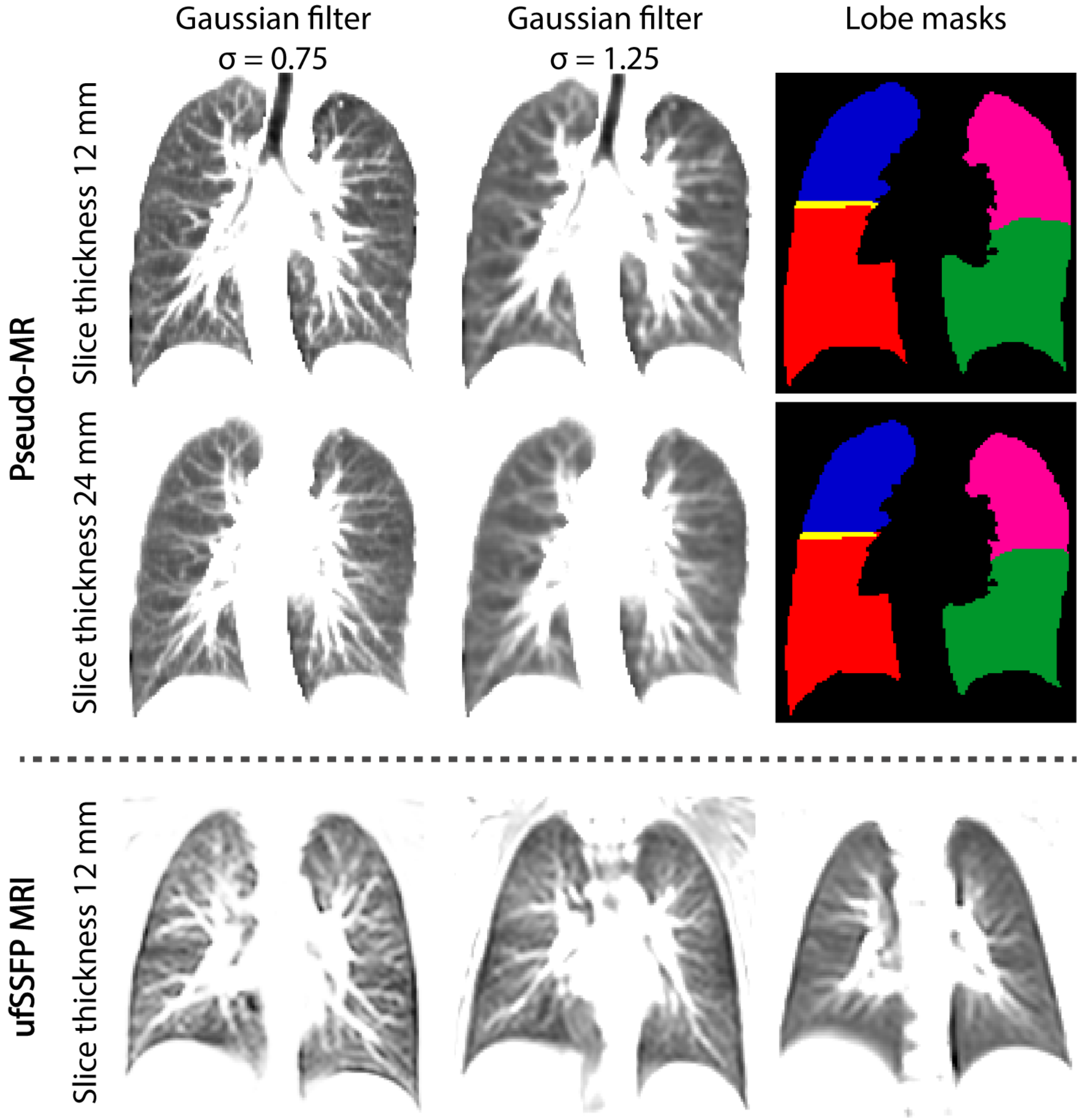

**Figure 4.** Pseudo-MR images reconstructed with 12 and 24 mm slice thickness (top and middle row, respectively) and Gaussian smoothing with σ = 0.75 and σ = 1.25, along with the respective lobe masks. In the bottom row, exemplary ufSSFP images in three subjects demonstrate the different degrees of visual sharpness present in the MR data, although the ufSSFP data are all acquired with the same resolution and slice thickness (12 mm). The generated pseudo-MR data appear visually similar to MR data (cf. vessel sharpness). To note, the pseudo-MR lobe segmentations are not identical (see the division between upper and lower lobes of the left lung) due to the different slice thicknesses and therefore partial-volume effects. The slice thickness influences only marginally the visual appearance of the peripheral lung vasculature of pseudo-MR images due to the simulated slice profile, which weights CT slices in the central part of the profile more than peripheral one (see subsection "CT data and lobe masks translation into pseudo-MR"). It has to be noted that this operation is very different from the well-known maximum intensity projection (see Figure 1), which would result in clearly visually different images.

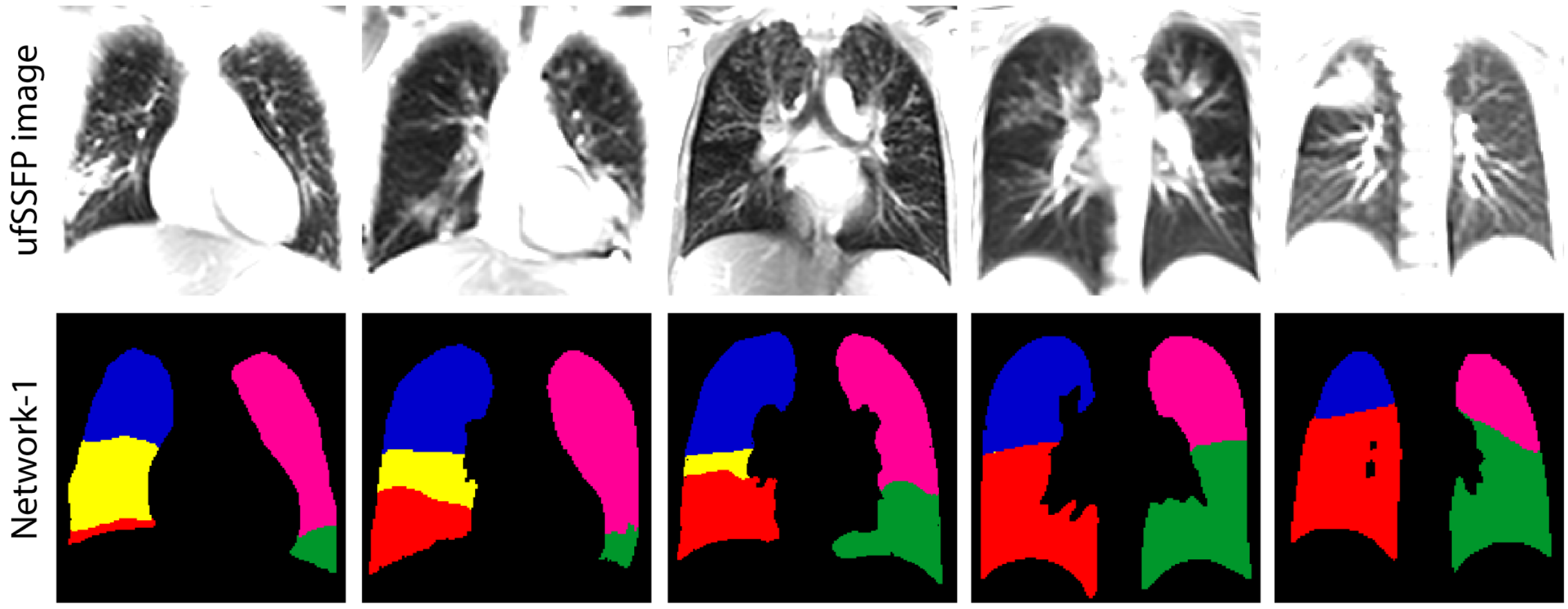

**Figure 5.** Lobe predictions of ufSSFP images with Network-1 in the CF cohort, which served as targets for the subsequent training of Network-2 and Network-3. Network-1 demonstrates ability to correctly predict lobes on MR images. The presented cases have been chosen and reordered to illustrate slices from anterior-to-posterior and, if possible, typical pathological CF lung characteristics. Additional examples are provided in Supporting Information Figure S2 available online.

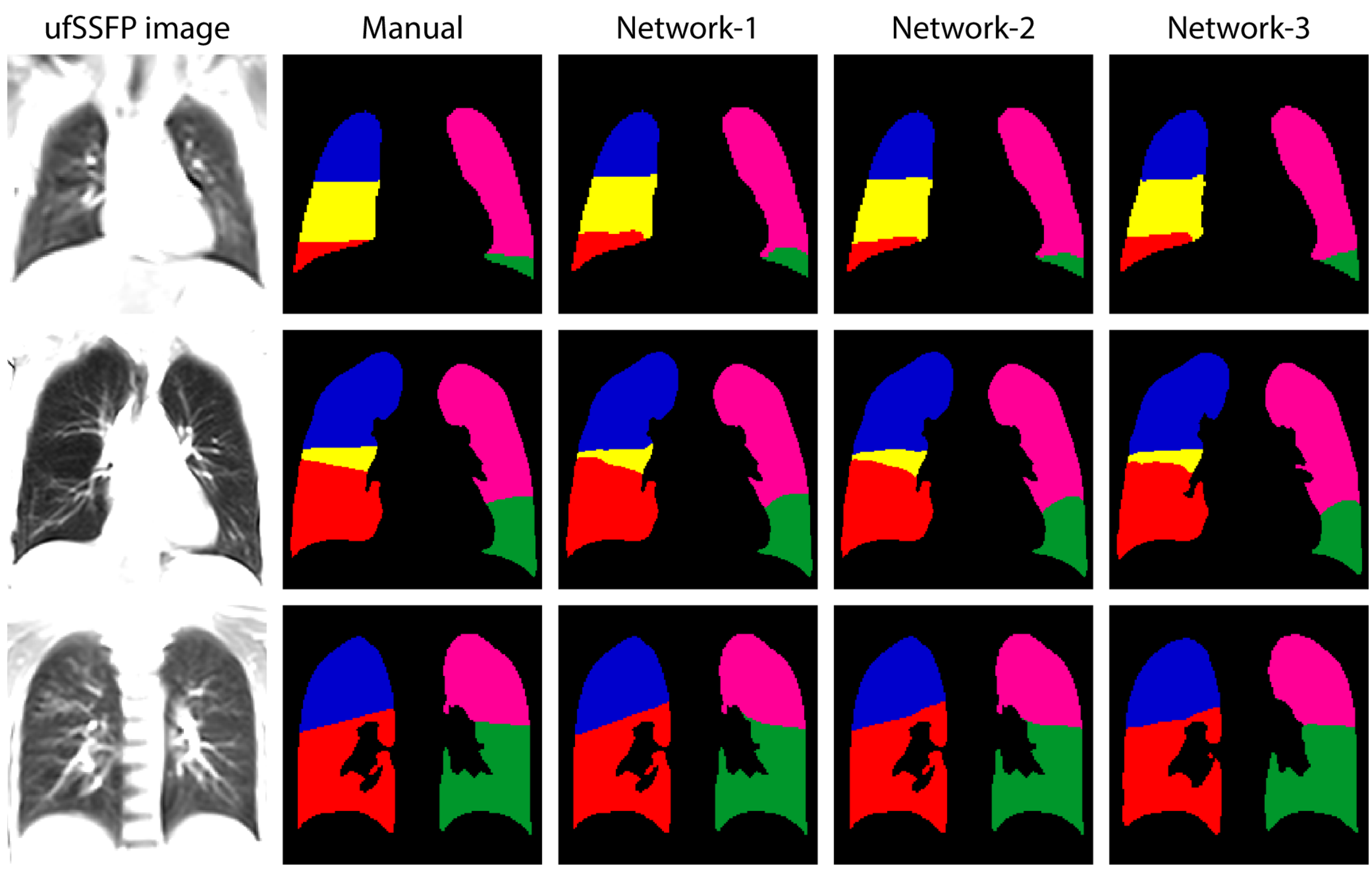

**Figure 6.** Representative ufSSFP images of five CF patients included in the testing group. Lung lobe segmentations were drawn manually and predicted with Network-1, Network-2, and Network-3. There is a good visual agreement among all the lobe segmentations. The differences between the segmentations are only marginal, e.g. note the shape of the right middle lobe masks in the second row.

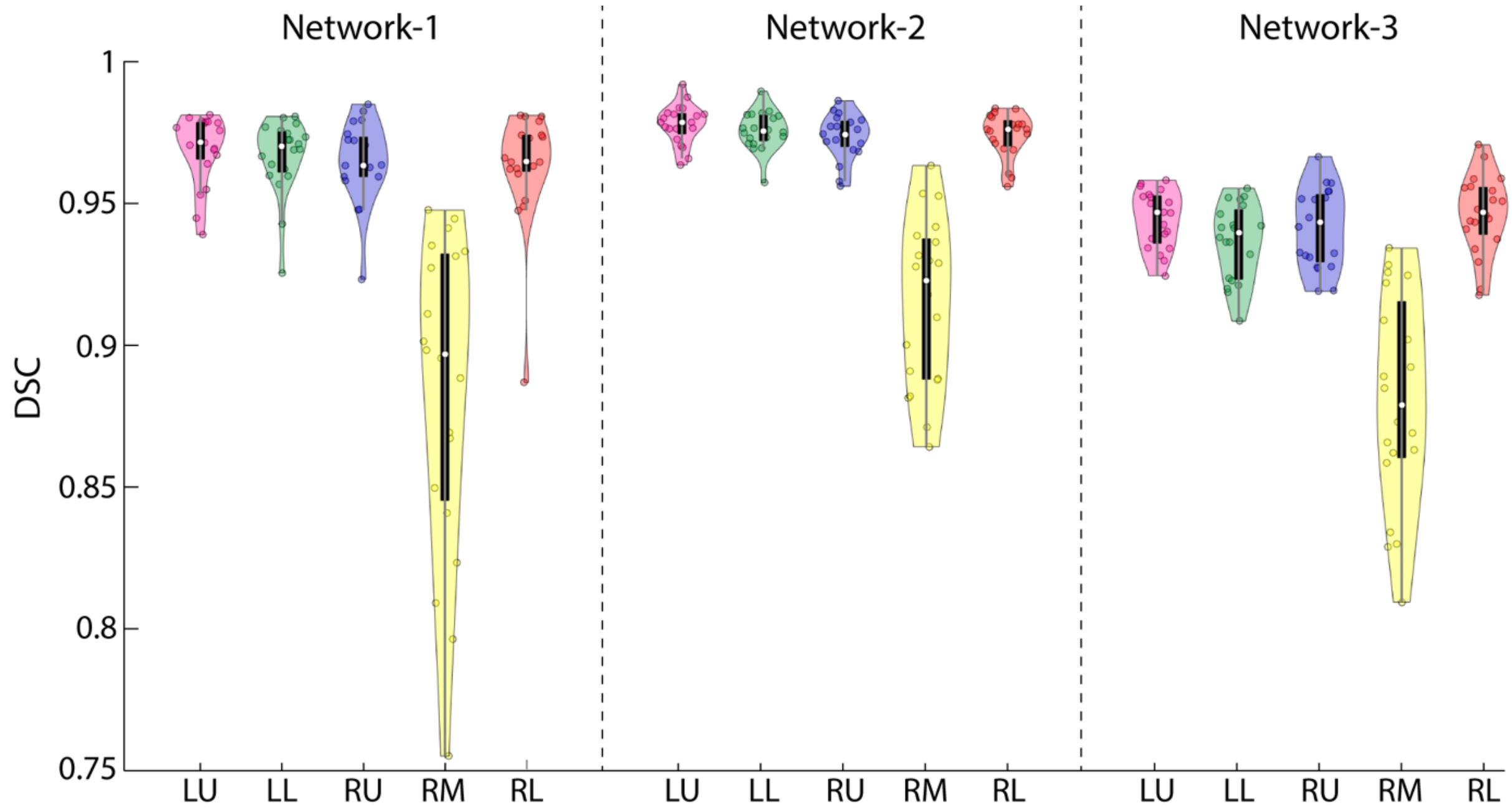

**Figure 7.** Violin plots reflecting the distribution of Dice similarity coefficients for the predictions of Network-1, Network-2, and Network-3 against the manual lobe segmentations. Data are evaluated in the MR testing CF cohort comprising 20 patients. The median values are displayed by the white dots. The thick black bars indicate the 25-75 percentile and the grey bars the whiskers. The right middle lobe reaches a lower DSC as compared to the other lobes due to its reduced size and the fact that DSC is sensitive to the relative size of the targets. To note, the HDs for the middle lobe presented in Table 1 do not indicate a lower accuracy as compared to the other lobes. Network-2 lobe predictions have a higher similarity to the manual segmentations, as compared to Network-1. The lower performance of Network-3 is caused by the lung boundaries which were not kept fixed. Abbreviations: LU= left upper lobe; LL=left lower lobe; RU=right upper lobe; RM=right middle lobe; RL=right lower lobe.

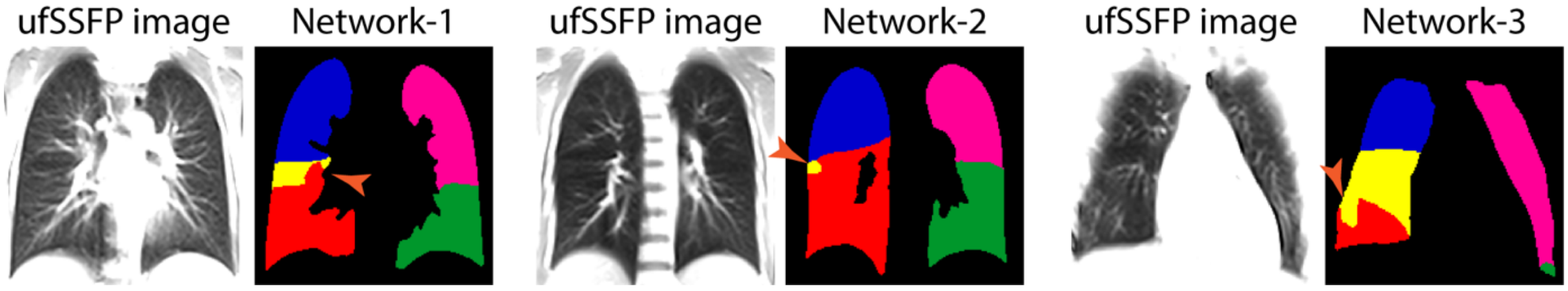

**Figure 8.** Examples of lobe segmentation flaws (orange arrowheads) that might occur with neural networks predictions. Data from 3 different CF subjects of the testing cohort are shown.

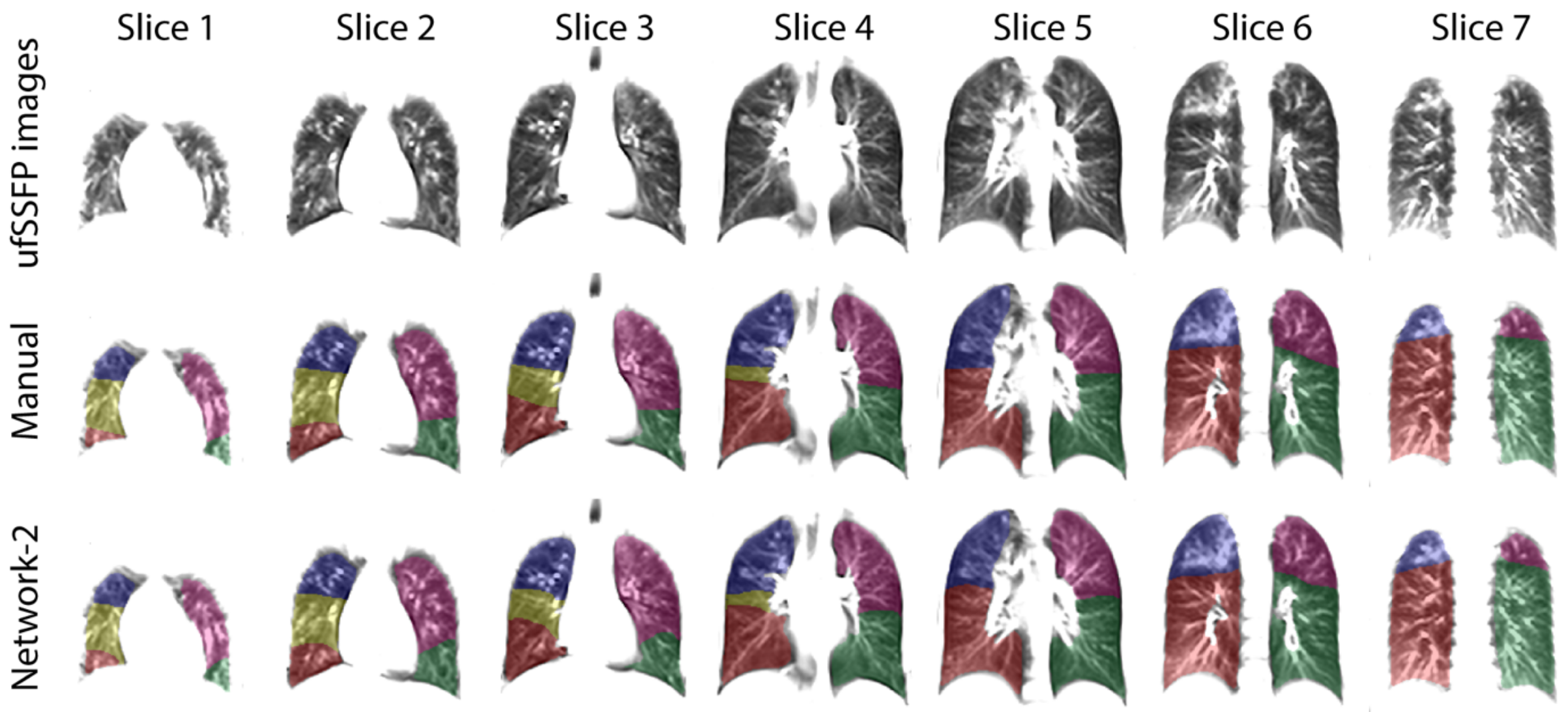

**Figure 9.** Whole-lung ufSSFP examination of a CF patient included in the testing cohort. The lobe masks are exemplarily displayed for manual segmentation and Network-2 overlaid on ufSSFP images. Mucus consolidations are visible, e.g., in the right upper lobe on slice 6. Network-2 segmented the lobes consistently among the slices, as observable in comparison to the manual segmentations. The lobe masks of the right lower lobe in slices 1 to 3 appear more consistent for the network prediction than for the manual drawn lobe masks. This corroborates that manual segmentation of ufSSFP data is very complex. Slice 5 was acquired just posterior to the pulmonary hila and thus no right middle lobe is present. In this patient, the average all-lobes DSC was 95.2±3.3 % (mean±SD) and the HD was 8.9±2.6 [mm] (mean±SD). The DSCs between manual lobe segmentations and Network-2 predictions were 97.6, 97.9, 94.7, 89.8, 96.0 [%] and the mean HDs were 8.7, 6.2, 6.7, 12.4, 10.7 [mm], for the left upper, left lower, right upper, right middle and right lower lobes, respectively. The corresponding lobe volume percentages were 22.3, 26.6, 15.6, 7.8, 27.8 [%] for the manual segmentations and 22.1, 26.8, 15.6, 7.9, 27.7 [%] for the segmentations derived with Network-2.

## Supporting Information

### Supporting Information Appendix S1

Input, target, and predicted data of Network-1, Network-2, and Network-3

• Network-1

The Network-1 was trained taking as input the pseudo-MR data (masked and normalized), and as target the reformatted pseudo-MR lung lobes (see Figure 2A and Figure 4). All input data for Network-1 are masked and normalized. To verify the lobe segmentation feasibility of 2D coronal imaging, Network-1 was used to predict the lung lobes of masked pseudo-MR testing data (cf. Figure 2), not included into training and validation sets.

The lung lobes were predicted by Network-1 for 1000 2D ufSSFP images (masked and normalized). These lobe segmentations served as a dictionary to train Network-2 and Network-3 (Figure 2B). Neither selection nor refinements of the data was performed to evaluate a workflow without user interaction. In a last step, Network-1 predicts the lung lobes of the CF testing cohort (Figure 2C), which are then compared to manually segmented lobe masks.

• Network-2

The Network-2 was trained taking as input both the ufSSFP data (non-masked and non-normalized) and the whole-lung segmentations derived with the network previously presented by Pusterla et al.[29], and as target the lung lobe masks predicted with Network-1 (Figure 2B). In the end, Network-2 predicts the lung lobes of the CF testing cohort, which are evaluated (Figure 2C).

• Network-3

The Network-3 was trained taking as input the ufSSFP data only (non-masked and non-normalized), and as target the lung lobe masks predicted with Network-1 (Figure 2B). Network-3 finally predicts the lung lobes of the CF testing cohort, which are evaluated (Figure 2C). Network-3 allows for an unbiased comparison to literature results since it does not rely on any a priori knowledge about the shape of the whole-lung (cf. Network-1 and Network-2 inputs).

To objectively compare the performance of Network-3 to the one of Network-1 and Network-2, the DSCs are also calculated by considering a fixed whole-lung boundary, i.e., lung lobe predictions of Network-3 are multiplied by the consensus whole-lung mask of Network-1 (which is the same as Network-2 and manual segmentations) before DSC computations.

**Supporting Information Table S1.** Dice similarity coefficients for lobe masks predicted with Network-1 on pseudo-MR data vs. ground truth. The pseudo-MR data were generated with 12, 18, and 24 mm slice thicknesses, and a Gaussian filter with σ = 0.75, σ = 1.0, and σ = 1.25.

| Slice thickness [mm] | σ Filter | | Left upper lobe | Left lower lobe | Right upper lobe | Right middle lobe | Right lower lobe | All-lobes mean (SD) |
|---|---|---|---|---|---|---|---|---|
| **12** | **0.75** | Mean | 98.4 | 97.9 | 97.1 | 91.9 | 97.7 | 96.6 (2.7) |
| | | SD | 1.4 | 2.4 | 1.3 | 4.7 | 1.2 | 2.2 |
| **12** | **1** | Mean | 98.4 | 97.9 | 97.1 | 91.9 | 97.7 | 96.6 (2.7) |
| | | SD | 1.4 | 2.5 | 1.4 | 4.5 | 1.1 | 2.2 |
| **12** | **1.25** | Mean | 98.3 | 97.8 | 97.0 | 91.8 | 97.7 | 96.5 (2.7) |
| | | SD | 1.4 | 2.5 | 1.4 | 4.2 | 1.1 | 2.1 |
| **18** | **0.75** | Mean | 98.2 | 97.5 | 97.3 | 91.7 | 97.7 | 96.5 (2.7) |
| | | SD | 1.5 | 2.7 | 1.4 | 4.2 | 1.3 | 2.2 |
| **18** | **1** | Mean | 98.1 | 97.5 | 97.3 | 91.8 | 97.8 | 96.5 (2.6) |
| | | SD | 1.6 | 2.8 | 1.3 | 3.5 | 1.3 | 2.1 |
| **18** | **1.25** | Mean | 98.0 | 97.5 | 97.3 | 91.7 | 97.7 | 96.4 (2.7) |
| | | SD | 1.6 | 2.8 | 1.2 | 3.4 | 1.3 | 2.1 |
| **24** | **0.75** | Mean | 98.1 | 97.3 | 97.4 | 91.4 | 97.5 | 96.3 (2.8) |
| | | SD | 1.2 | 3.0 | 1.2 | 4.3 | 1.3 | 2.2 |
| **24** | **1** | Mean | 98.1 | 97.3 | 97.4 | 91.4 | 97.5 | 96.3 (2.8) |
| | | SD | 1.4 | 3.5 | 1.1 | 4.1 | 1.4 | 2.3 |
| **24** | **1.25** | Mean | 98.0 | 97.1 | 97.3 | 91.1 | 97.4 | 96.2 (2.9) |
| | | SD | 1.7 | 4.0 | 1.2 | 4.3 | 1.4 | 2.5 |

**Supporting Information Figure S1**

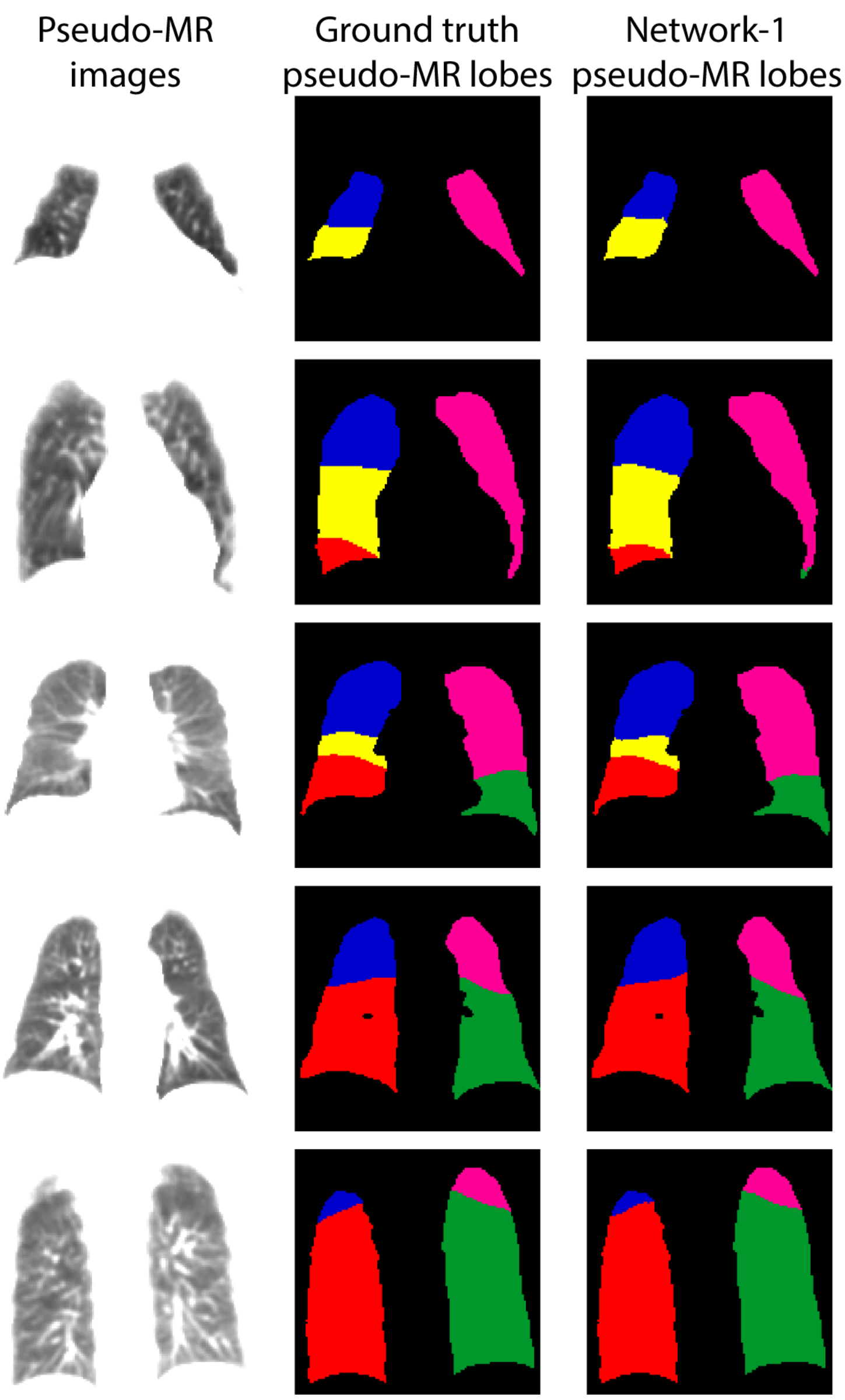

**Figure S1.** Pseudo-MR images as well as ground truth lobes and predictions with Network-1. The lung lobe predictions are feasible over the whole lung and appear well performed, as corroborated by the DSCs. The network predictions seem less accurate in very anterior slices, as observable in the first image, on the top. Pseudo-MR are displayed for an 18 mm slice thickness and a Gaussian filter with $\sigma$=1.

**Supporting Information Figure S2**

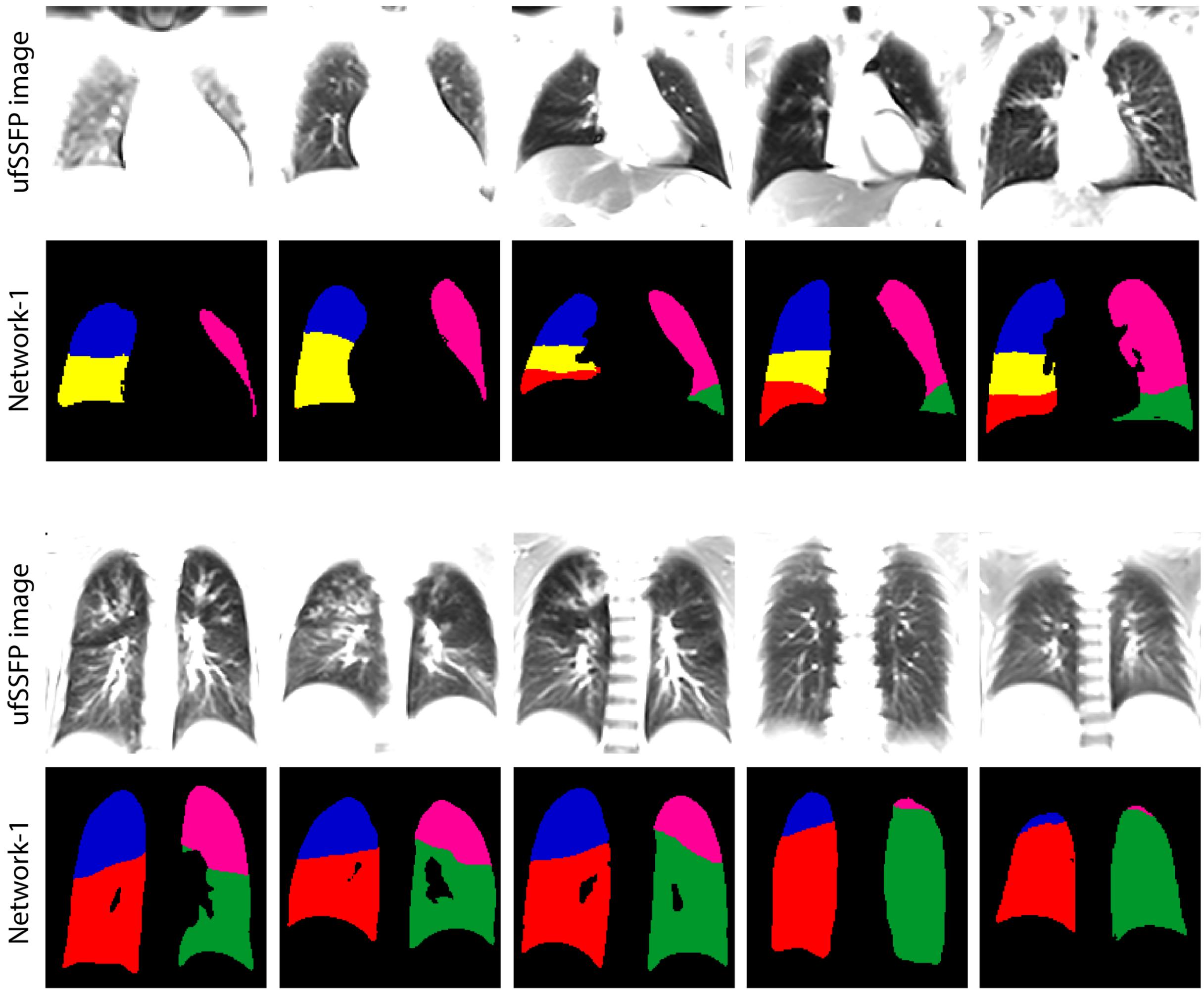

**Figure S2.** Additional lobe prediction examples predicted with Network-1 in the CF cohort (similarly to Figure 5 in the manuscript). The lobe segmentations appear correctly predicted. Network-1 lobe predictions served as targets for the subsequent training of Network-2 and Network-3. The presented cases have been chosen and reordered to illustrate slices from anterior-to-posterior and, when possible, typical pathological CF lung characteristics.

**Supporting Information Figure S3**

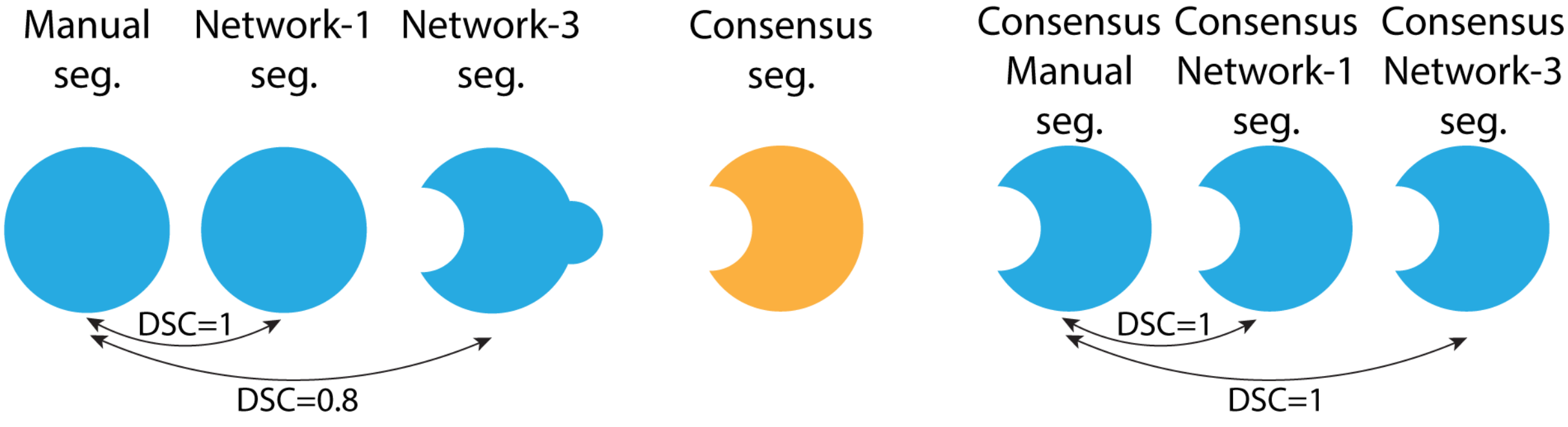

**Figure S3.** The manual whole-lung segmentation and the one derived with Network-1 (and Network-2, not shown) are exactly the same, while the one of Network-3 is different in shape (left). The DSCs are calculated using the manual segmentation as reference. A consensus segmentation (middle) is computed (Manual seg. ∩ Network-3 seg.) and applied to all the segmentations (right; e.g., Consensus seg. ∩ Manual seg., Consensus seg. ∩ Network-1 seg., etcetera). Based on the consensus segmentation, the DSCs are recalculated. Only the DSC computed between the consensus manual segmentation and consensus Network-3 segmentation is modified and improved.